\documentclass[letter]{aa}  

\usepackage{graphicx}
\usepackage{txfonts}

\begin{document}

   \title{Multiple retrograde substructures in the Galactic halo: \\ A shattered view of Galactic history}
  \titlerunning{Multiple retrograde substructures in the Galactic halo}
   
   \author{Helmer H. Koppelman \inst{1}
          \and Amina Helmi \inst{1}
          \and Davide Massari \inst{1,2,3}
          \and Adrian M. Price-Whelan \inst{4}
          \and Tjitske K. Starkenburg \inst{4}
          }

        \institute{Kapteyn Astronomical Institute, University of Groningen, Landleven 12, 9747 AD Groningen, The Netherlands\\
              \email{koppelman@astro.rug.nl}
        \and Dipartimento di Fisica e Astronomia, Universit\`{a} degli Studi di Bologna, Via Gobetti 93/2, I-40129 Bologna, Italy
        \and INAF - Osservatorio di Astrofisica e Scienza dello Spazio di Bologna, Via Gobetti 93/3, I-40129 Bologna, Italy
        \and Center for Computational Astrophysics, Flatiron Institute, 162 5th Avenue, New York, NY 10010, USA
}
   \date{}

  \abstract
   {}
   {
   Several kinematic and chemical substructures have been recently found amongst Milky Way halo stars with retrograde motions. It is currently unclear how these various structures are related to each other. This Letter aims to shed light on this issue.
   }
   {
   We explore the retrograde halo with an augmented version of the {\it Gaia} DR2 RVS sample, extended with data from three large spectroscopic surveys, namely RAVE, APOGEE and LAMOST. In this dataset, we identify several structures using the {\tt HDBSCAN} clustering algorithm. We discuss their properties and possible links using all the available chemical and dynamical information.
   }
   {
   In concordance with previous work, we find that stars with [Fe/H]~$ < -1$ have more retrograde motions than those with [Fe/H]~$ > -1$. The retrograde halo contains a mixture of debris from objects like Gaia-Enceladus, Sequoia, and even the chemically defined thick-disc. We find that the Sequoia has a smaller range in orbital energies than previously suggested and is confined to high-energy. Sequoia could be a small galaxy in itself, but since it overlaps both in integrals-of-motion space and chemical abundance space with the less bound debris of Gaia-Enceladus, its nature cannot be fully settled yet. In the low-energy part of the halo we find evidence for at least one more distinct structure: {\it Thamnos}. Stars in Thamnos are on low inclination, mildly eccentric retrograde orbits, moving at $v_{\phi}\approx-150~{\rm km/s}$, and are chemically distinct from the other structures.
   }
   {
   Even with the excellent {\it Gaia} DR2 data it remains challenging to piece together all the fragments found in the retrograde halo. 
   At this point, we are very much in need of large datasets with high-quality high-resolution spectra and tailored high-resolution hydrodynamical simulations of galaxy mergers.
   }


   \maketitle
%

\section{Introduction}
A wide variety of cosmological simulations, typically performed in a
$\Lambda$CDM setting, have shown that the stellar halo of the Milky
Way is an excellent testbed for galaxy formation models
\citep{Helmi2003TheHalo,Bullock2005,Johnston2008,Cooper2010,
  Pillepich2014, Grand2017TheTime}. In $\Lambda$CDM, the halos of
galaxies like the Milky Way grow in size by merging with other
galaxies, mostly through minor mergers. Galaxies that merge leave behind debris in the form of 
a trail of stars, and at the solar position this debris typically is very phase-mixed
\citep{Helmi1999a}. Disentangling the superimposed trails of different
mergers is in principle possible with the help of detailed
dynamical information like the integrals of motion \citep{Helmi2000},
or the actions \citep{McMillan2008}. In a local volume, each stream 
(a portion of a trail with stars with similar
orbital phase) has typically a very low density, and has been estimated to contain on average 0.25\% and at maximum 5\% of the total number of local halo stars \citep{Gould2003ANHALO}. 

The Milky Way's outer stellar halo is consistent with being build-up
fully through mergers \citep[e.g.][]{Belokurov2006,Bell2008,Helmi2011SUBSTRUCTURESIMULATIONS}. 
With {\it Gaia} \citep{GaiaCollaboration2016TheMission, GaiaCollaboration2018brown} it
has become possible to map the kinematics of the local stellar halo in great detail
\citep[e.g.][]{Helmi2017,Myeong2018c,Myeong2018,Koppelman2018}. An
impressive finding in the field of Galactic archaeology since the
release of {\it Gaia} DR2 is the debris of Gaia-Enceladus-Sausage
\citep{Belokurov2018Co-formationHalo,Helmi2018}: a massive dwarf
galaxy that contributed a large fraction of the local stellar
halo. This object's initial stellar mass was 
$5\cdot10^{8} - 5\cdot10^{9}~{\rm M}_\odot$ \citep{Belokurov2018Co-formationHalo, Helmi2018,
  Mackereth2019TheSimulations,Vincenzo2019TheSausage} and it was accreted
$\sim 10$ Gyr ago \citep{Helmi2018, DiMatteo2018,  Gallart2019UncoveringGaia}.

Besides Gaia-Enceladus, the Helmi streams \citep{Helmi1999} are located
 in the prograde part of the halo. These streams originate in a
dwarf galaxy of $M_\star \sim 10^8~{\rm M}_\odot$ that was accreted
5-8 Gyr ago \citep[][see also
\citealt{Kepley2007HALONEIGHBORHOOD}]{Koppelman2019CharacterizationDR2}. While
a large fraction of stars with retrograde motions appears to be debris
from Gaia-Enceladus especially for high-eccentricity \citep[see also
\citealt{Belokurov2018Co-formationHalo}]{Helmi2018}, for very retrograde
motions ($v_\phi < -100$~km/s) the situation is less clear. This
portion of the halo contains several small structures
\citep[e.g.][]{Myeong2018, Koppelman2018, Matsuno2019OriginHalo}, and
plausibly also debris of Gaia-Enceladus. Also
\cite{Mackereth2019TheSimulations} postulate that the low-eccentricity
region had a more complex formation history and would be composed by
a mixture of stars formed {\it in situ}, debris from Gaia-Enceladus,
and debris from other structures. One such structure would be the
Sequoia \citep{Myeong2019EvidenceHalo}, whose existence builds on the
discovery of a large globular cluster with very retrograde halo-like
motion, FSR-1758 \citep{Barba2019ACluster}.

In this {\it Letter} we quantify the degree of clustering in a local
sample of halo stars using both dynamical and metallicity
information. This allows us to discover debris from another small
object, which we term Thamnos, as well as to establish on firmer
grounds the reality and relationship between the different structures
reported thus far in the literature in this rapidly evolving field.

\section{Data}\label{sec:data}
We use here an augmented version of the {\it Gaia} RVS sample, 
extended with radial velocities from APOGEE DR14 
\citep{Abolfathi2018TheExperiment}, LAMOST \citep{Cui2012}, 
and RAVE DR5 \citep{Kunder2017}, see Sect.~2.1 and 2.2 of 
\cite{Koppelman2019CharacterizationDR2} for more details. 
Because the metallicity scales of the three different surveys 
are not necessarily the same, we will use the LAMOST 
values, unless mentioned otherwise. The results do
not depend on this choice, except 
that the cross-matches with APOGEE and RAVE have considerable fewer stars. 
In total, our sample comprises $8~738~322$ stars 
with full 6D phase-space information and high-quality parallaxes (${\tt parallax\_over\_error>5}$) of which $3~404~432$ have 
additional [Fe/H] information and $189~444$ have chemical 
abundances from APOGEE. To calculate the distance we invert 
the parallaxes. Because of the systematic parallax offset in 
{\it Gaia} DR2 \citep{Arenou2018,GaiaCollaboration2018brown,
Lindegren2018}, which for the RVS sample might even be more 
significant \citep{Schonrich2019DistancesDR2}, we restrict 
our analysis to stars within 3 kpc of the Sun. When inspecting 
velocities we use a selection of stars in an even smaller volume to 
optimise the amount of clumpiness (by avoiding possible velocity 
gradients).

The velocities of the stars are corrected for the solar motion 
assuming $(U,V,W) = (11.1,12.24,7.25)~{\rm km/s}$ \citep{Schonrich2010}, 
and for the motion of the LSR using $v_{LSR} = 232.8~{\rm km/s}$ 
\citep{Mcmillan2017}. Cartesian coordinates are calculated 
such that $X$ points towards the Galactic Centre, and $Y$ 
points in the direction of the motion of the disc. Cylindrical 
coordinates are derived in a right-handed system, although we flip 
the sign of $v_\phi$ such that it coincides with the $Y$-axis 
at the solar position. In this system, the Sun is located at 
$X=-8.2~{\rm kpc}$. We use the implementation of the 
\cite{Mcmillan2017} potential in {\tt AGAMA} \citep{Vasiliev2019} 
to calculate orbital parameters such as the total energy $(En)$, 
eccentricity $(ecc)$, circularity $(circ)$, apocentre $(apo)$, 
and pericentre $(peri)$. The circularity is calculated as 
$circ = L_z/|L_{z,circ}|$, where $L_{z,circ}$ is the vertical 
component of the angular momentum for a circular orbit with 
the same $En$ of the star.

In this work, we identify halo stars by their kinematics, a 
selection mostly used for illustrative purposes. As we are mainly 
interested in the retrograde halo we impose a relatively 
conservative cut by removing stars with $|{\bf V}-{\bf V}_{LSR}|<230~{\rm km/s}$. 
\section{Results}

\subsection{The metal-poor, retrograde halo}
\begin{figure}
    \centering
    \includegraphics[width=\hsize]{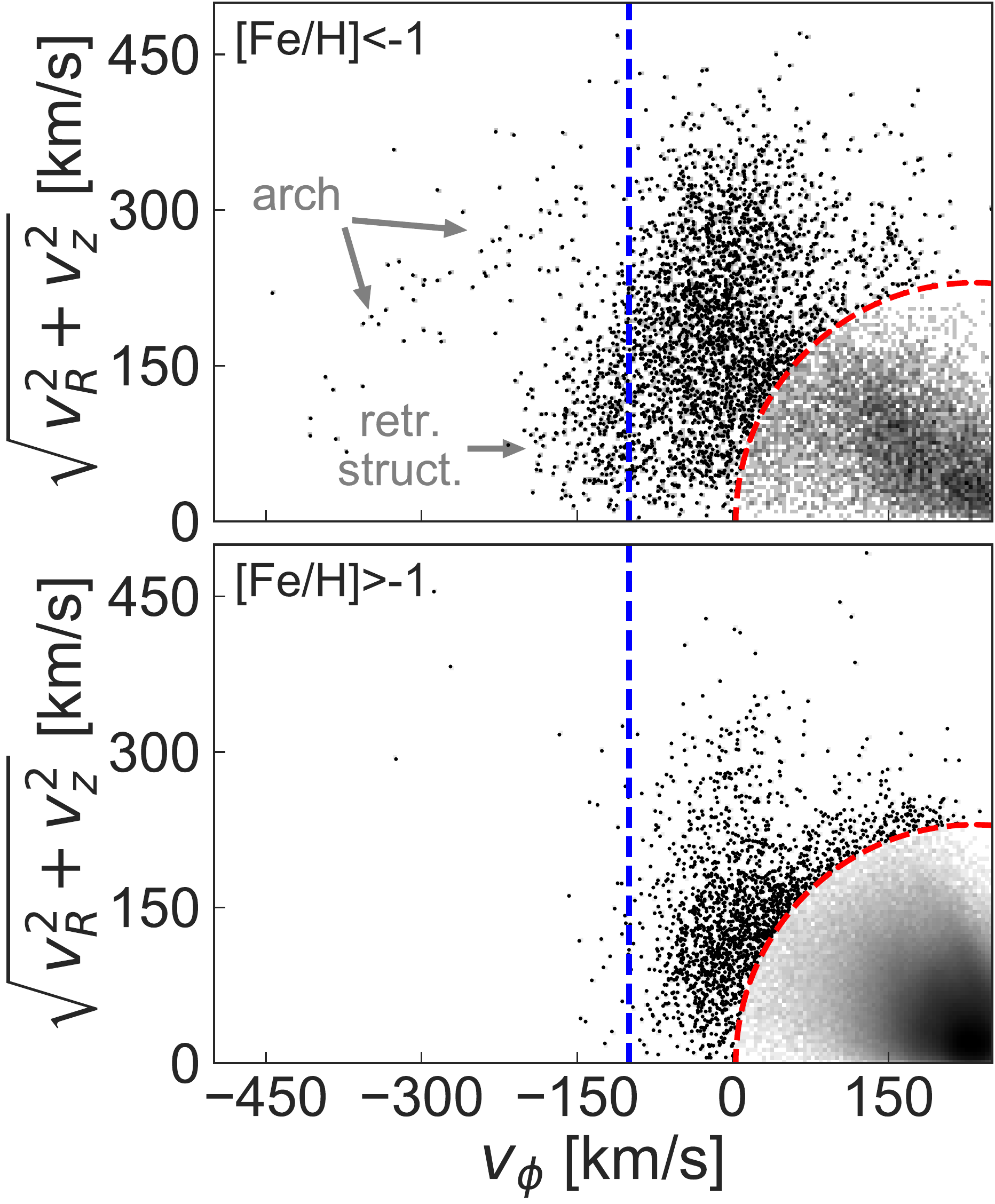}
    \caption{Velocity diagram of the local $(d<1~{\rm kpc})$ stellar halo split
      in a metal-poor (top) and a metal-rich sample (bottom). The
      colour-coding of the 2D histogram scales with the logarithm of
      the number of stars in each bin. All stars outside of the red
      dashed line are tentatively labelled as halo stars and are
      shown as black dots. Note that most, if not all
      of the halo left of the dashed
      vertical line $(v_\phi < -100~{\rm km/s})$ is more metal-poor 
      than ${\rm [Fe/H]}=-1$.}
    \label{fig:vel-feh-cut}
\end{figure}

Figure~\ref{fig:vel-feh-cut} shows a velocity diagram of the local 
stellar halo (distance $<1$ kpc) split in a metal-poor (top) and a 
metal-rich (bottom) sample. 2D-histograms show the distribution of 
all the stars in the given [Fe/H] selection, while halo stars are 
highlighted with small black dots. The vertical dashed line indicates 
the very retrograde limit 
and highlights the large amount of small-scale substructure present 
for low metallicity. This is consistent with previous work reporting 
that the retrograde halo is more metal-poor \citep[e.g.][]{Carollo2007,
Matsuno2019OriginHalo,Myeong2019EvidenceHalo}. One of the structures 
seen is the arch reaching from $(v_\phi, (v_z^2+v_R^2)^{1/2}) = (-100,300)$ 
to $(-450,0)$ km/s, which overlaps with the retrograde structures 
of \cite{Myeong2018} and with the red and purple structures in 
\cite{Koppelman2018}. The arch was associated to Gaia-Enceladus \citep{Helmi2018} on 
the basis of resemblance to the simulations of \cite{Villalobos2008}. 

Besides the arch, there is another retrograde structure apparent in
the metal-poor halo, at $v_\phi = - 150~{\rm km/s}$ and with
$(v_z^2+v_R^2)^{1/2} < 150~{\rm km/s}$, i.e. with 
counter-rotating thick-disc-like kinematics. A subset of this retrograde
component was picked up as the VelHel-4 structure \cite{Helmi2017} and
as the blue and orange structures reported in \cite{Koppelman2018}.

The debris of Gaia-Enceladus, which we identify here as the dominant
contributor to the halo in the range $-100 < v_\phi < 50$~km/s, has more stars with [Fe/H]~$< -1$ (top panel) but also contributes to the metal-rich (bottom) panel. The only structure that is more abundant in the
metal-rich part of the halo is the extension of the thick-disc, 
identified as the slow-rotating tail of the thick disc
\citep[e.g.][]{Koppelman2018,Haywood2018InDR2,DiMatteo2018}. Stars with thin-disc-like motions appear to also exist with [Fe/H]~$< -1$.

\subsection{Selecting distinct substructures}
\begin{figure*}
    \centering
    \includegraphics[width=\hsize]{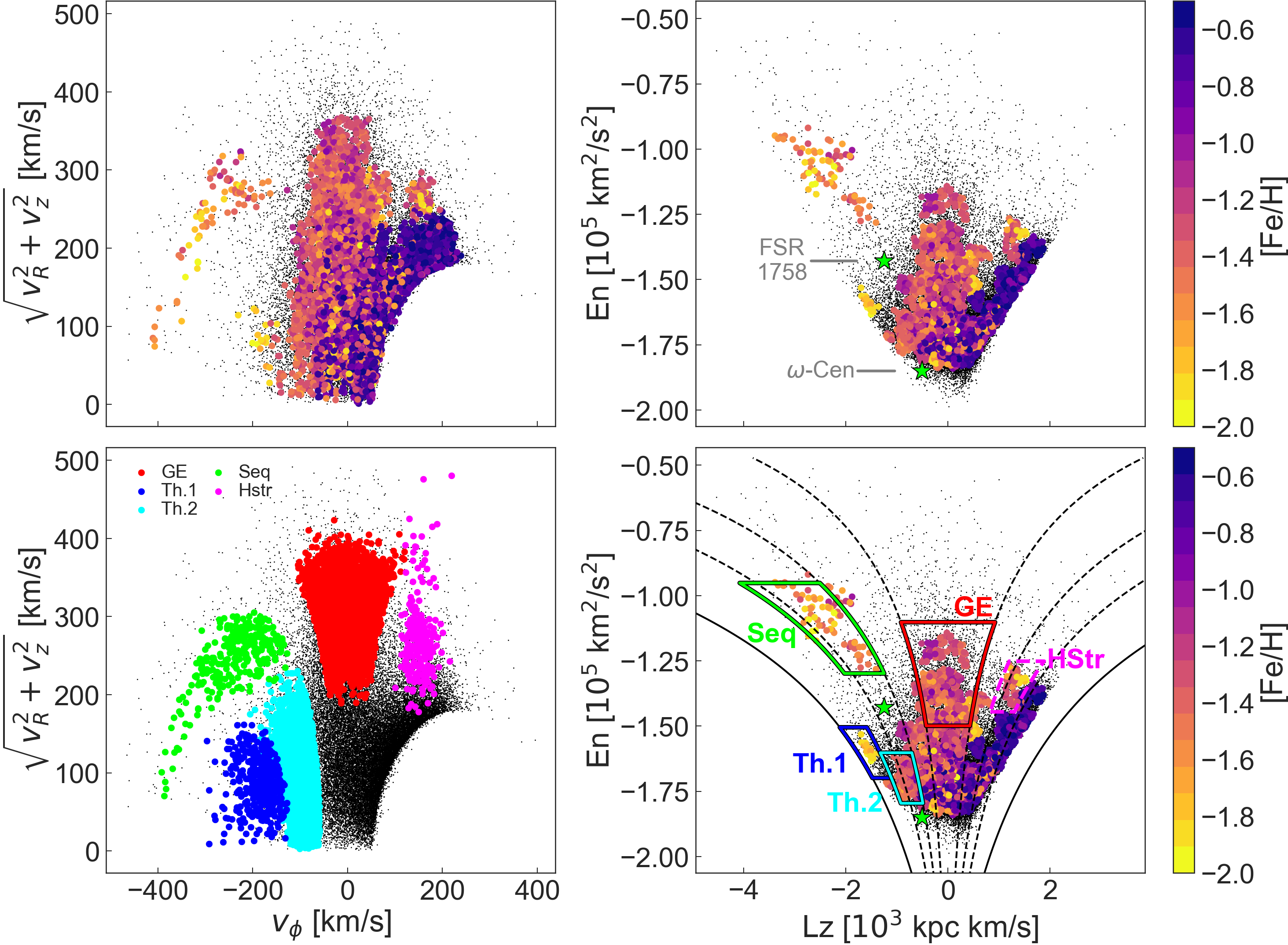}
    \caption{
    The top rows show the distribution of the stars in the groups identified by HDBSCAN in En, Lz, ecc, and [Fe/H] space, and colour-coded by [Fe/H], with the rest shown with black dots. In the bottom-right panel, we have over-plotted lines of constant circularity and used coloured boxes to indicate our selection of substructures. Note that the Helmi Streams (HStr) are selected in $L_z - L_\perp$ space as described in the text. The bottom-left panel shows the kinematic properties of the stars in these substructures.}
    \label{fig:clustering-results}
\end{figure*}
\begin{figure*}
    \centering
    \includegraphics[width=\hsize]{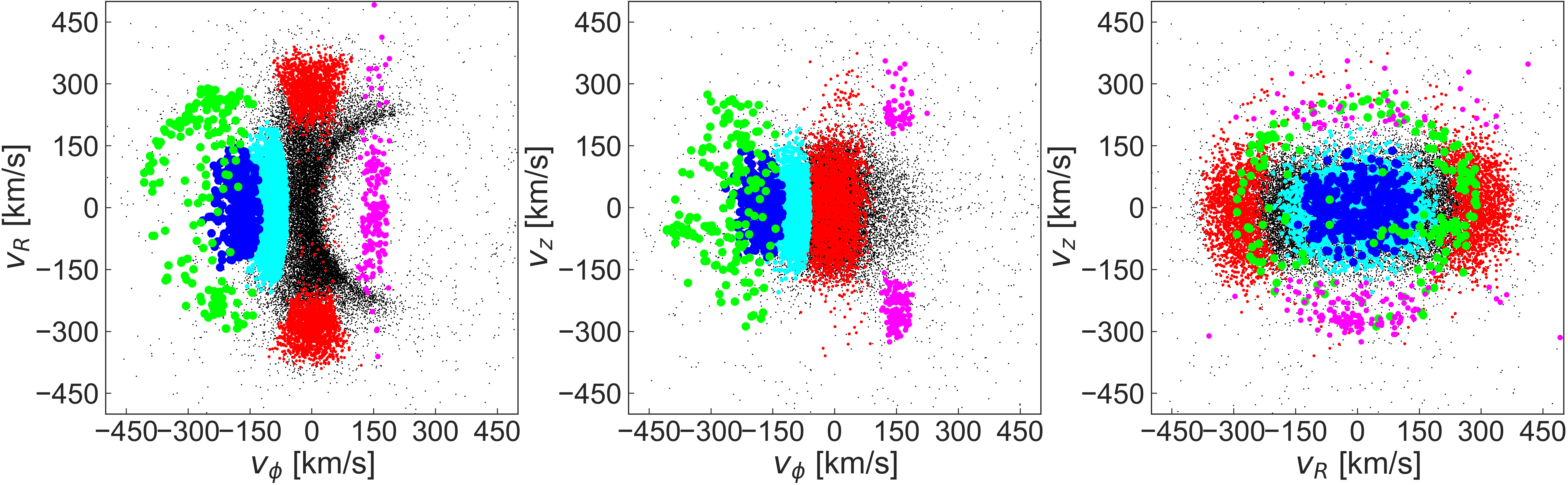}
    \caption{Velocity distributions of the stars in the structures identified in Fig.~\ref{fig:clustering-results} (using the same colour-coding), with non-selected stars with halo-like kinematics stars in black. Only stars within 2 kpc are shown here to optimise the amount of clumpiness in this space (by minimising velocity gradients).}
    \label{fig:finalvelplot}
\end{figure*}

Figure~\ref{fig:vel-feh-cut} on its own does not clear up if and how the retrograde structures are related. To study this in more detail we apply the clustering algorithm {\tt HDBSCAN}\footnote{Hierarchical Density-Based Spatial Clustering of Applications with Noise, a clustering algorithm that excels over the better known {\tt DBSCAN} both because it is less sensitive to the parameter selection and because it can find clusters of varying densities. See also https://hdbscan.readthedocs.io}\citep{McInnes2017Hdbscan:Clustering}. We use the algorithm's default parameters, after setting ${\tt min\_samples} = 3$, ${\tt min\_cluster\_size} = 15$, and ${\tt cluster\_selection\_method}=${\tt`leaf'}. These settings, especially the {\tt leaf} mode, tune the algorithm to find fine-grained structure instead of large overdensities. In our experience, no clustering algorithm picks out halo overdensities uniquely given the large amount of overlap, the measurement errors, and the lack of metallicities for most sources. Therefore, we aim to break up the halo in small, robust groups that can be used to trace the large structures. Based on these groups we then place selection boxes to select the larger structures.

As input-parameters for the algorithm we use $En$, $L_z$, $ecc$,
and [Fe/H], which are all often used to find substructure in the
stellar halo
\citep[e.g.][]{Helmi2000,Helmi2017,Koppelman2018,Mackereth2019TheSimulations}. The
space that is defined by these parameters is scaled with {\tt
  RobustScaler} implemented in {\tt scikit-learn}
\citep{Pedregosa2011Scikit-learn:Perrot} using the code's default
settings. We select all stars within 3 kpc of the Sun and
$|{\bf V}-{\bf V}_{LSR}|>180~{\rm km/s}$, because we are mainly
interested in picking up structure in the halo. This selection
includes a significant amount of thick-disc stars that should be identified
as a distinct component if the algorithm works properly. 
There are no thin-disc stars in this selection.

Figure~\ref{fig:clustering-results} shows the stars associated with
substructures according to {\tt HDBSCAN}, colour-coded by [Fe/H], while the
remaining stars are shown with black dots. The top-left panel is
similar to Fig.~\ref{fig:vel-feh-cut} and shows a very clear gradient
of metallicity with $v_\phi$. Both the arch and the low
$(v_z^2+v_R^2)^{1/2}$ structures are picked up as (metal-poor) groups
(in yellow), while the thick disk is apparent (in purple) too. When
varying the {\tt HDBSCAN} parameters the individual groups change
slightly, but the large structures which they trace persist. The
results are also robust to changes in the limiting distance of the
stars, at least up to 5 kpc from the Sun.

The top-right panel of Fig.~\ref{fig:clustering-results} shows the
distribution of the clusters in $En$-$L_z$ space. In this space, it
becomes clear that the arch structure strongly overlaps with the retrograde
group (i.e. Sequoia) identified by \cite{Myeong2018}. As a reference we add the globular clusters FSR 1758 and $\omega$-Cen to this diagram, both of which have 
tentatively been assigned to the Sequoia by \citet[][although \citet{Massari2019TheClusters} argues that the latter is more likely associated with Gaia-Enceladus]{Myeong2019EvidenceHalo}.
In the
bottom-right panel of Fig.~\ref{fig:clustering-results} we have
overlaid lines of constant circularity ($circ$ = $-0.2,-0.4,-0.6$),
with solid lines corresponding to circular orbits in the Galactic
plane. We select here regions occupied {\it predominantly} by the various
structures as follows:
\begin{itemize}
    \item Gaia-Enceladus: $-1.5<En/[10^{5}~{\rm km^2/s^2}]<-1.1$
and $|circ|<0.2$;
    \item Sequoia: $-1.35<En/[10^{5}~{\rm km^2/s^2}]<-1.0$ and 
$-0.65<circ<-0.4$; 
    \item The low $(v_z^2+v_R^2)^{1/2}$ structure is split in two
based on the different metallicities: 
$i)$: with
$v_\phi\sim-200$~km/s, $circ<-0.75$ and
$-1.65<En/[10^{5}~{\rm km^2/s^2}]<-1.45$; and \\$ii)$: with 
$v_\phi\sim-150$~km/s, $-0.75<circ<-0.4$ and
$-1.8<En/[10^{5}~{\rm km^2/s^2}]<-1.6$. 
\end{itemize} 
We show with different colours, how the stars in these selections are distributed in velocity space in the bottom-left panel of Fig.~\ref{fig:clustering-results}. As a reference, we also plot the Helmi streams (HStr), selected as all stars with $1600<L_\perp / {\rm [kpc~km/s]}<3200$ and $1000<L_z /{\rm [kpc~km/s]} <1500$ \cite[c.f.][]{Koppelman2019CharacterizationDR2}. Note that in this figure, no globular clusters are found on the region occupied by the stars colour-coded dark blue \citep[c.f.][]{Massari2019TheClusters}. On the other hand, the cyan stars are located near $\omega$-Cen and hence if we follow the argument of  \cite{Myeong2018}, they could belong to the Sequoia. 
It is possible however that these cyan stars are tracing a new structure, or are associated with those in the dark blue selection, or be a mixture of things. 

For completeness, in Figure~\ref{fig:finalvelplot} we plot the structures in other projections of velocity space. Their distribution is very reminiscent of the simulated substructures in \cite{Helmi2000}, suggesting that they could indeed belong to different dwarf-galaxy progenitors.

\subsection{Chemical analysis}
\begin{figure}
    \centering
    \includegraphics[width=\hsize]{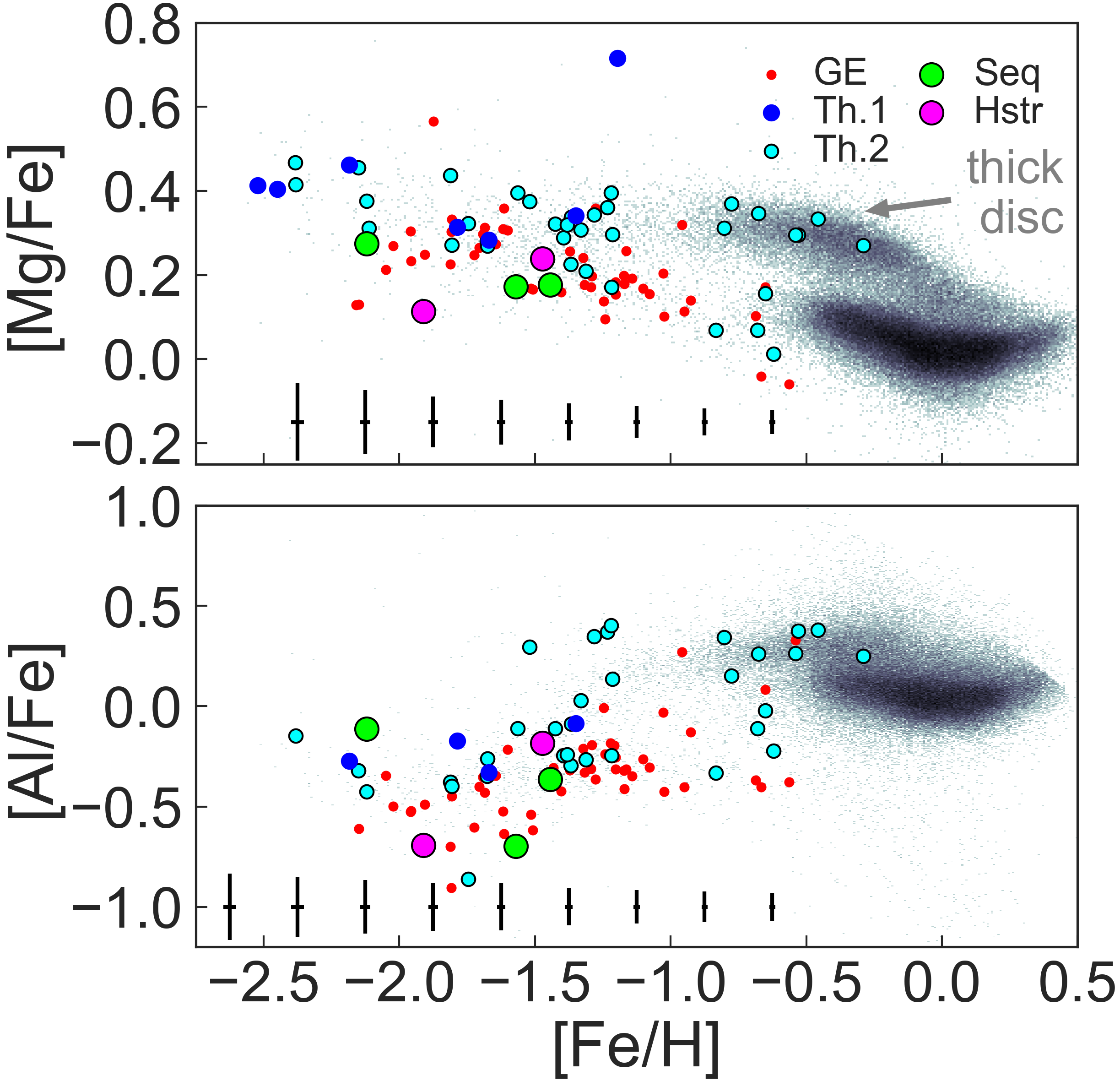}
    \caption{Detailed chemical abundances (from APOGEE) for stars in the 
      different substructures. The bars at the bottom of the
      panel indicate the mean error at that [Fe/H]. In the background
      we show a 2D histogram of all of the stars in our dataset,
      colour-coded by the logarithm of the number of stars per bin.}
    \label{fig:apogee}
\end{figure}

Figure \ref{fig:apogee} shows the distribution of stars in our sample
with abundances from APOGEE (with ${\tt ASPCAPFLAG}==0$), colour coded
according to our selections. Especially for [Fe/H]~$> -1.5$, we see
that the stars in cyan reveal contamination from the
chemically defined thick-disc as indicated by the annotation (despite their very retrograde motion) and
Gaia-Enceladus. For low [Fe/H], these stars typically have higher
${\rm [Mg/Fe]}$ than Sequoia and Gaia-Enceladus, indicating a
different origin. We tentatively refer to the structure defined by
the cyan and dark blue stars as {\it Thamnos}, i.e. ``shrubs'', because these stars
stand at the foot of a Greek giant and a tall tree in both velocity
and $En-L_z$ spaces. We keep for now the distinction between the stars
with $v_\phi\sim-200$~km/s and those with $v_\phi\sim-150$~km/s and
refer to them as {\it Thamnos 1} and {\it 2}, respectively.

The Sequoia stars, on the other hand, overlap with the metal-poor tail
of Gaia-Enceladus, making it difficult to argue that they truly
originate in a different system. It should be noted that much
of this analysis is tentative as it is only based on a small sample of
stars and at low [Fe/H] the errors are significant. Furthermore, the
various other (independent) elements in APOGEE also have too large
errors to be of help. With the amounts of data coming in the next few years 
this analysis will be much improved. 
\section{Discussion and Conclusions}

We have used {\it Gaia} DR2 data, supplemented with line-of-sight
velocities and chemical abundances from RAVE, APOGEE and LAMOST 
to shed more light on the nearby stellar halo and its
substructures. The stars in the retrograde halo are predominantly metal-poor. In fact, as the stars' motions become more retrograde, the stars are even more metal-poor (i.e. a ``gradient'' in $v_\phi$ with [Fe/H]). 
This gradient is reminiscent of the dual halo reported
in \cite{Carollo2007}, but its nature is more complex. Our analysis seems
to suggest that the outer halo is more retrograde because it is
dominated by debris from (the outskirts of) Gaia-Enceladus and
Sequoia. This was already hinted at by \cite{Helmi2017}, who have shown 
that at high energies the halo is retrograde. On the other hand, stars 
on very retrograde motions with orbits in the inner halo belong to a 
newly-identified \citep[but previously reported in part in][]{Helmi2017,
Koppelman2018,Mackereth2019TheSimulations} substructure which we have 
named {\it Thamnos}.

\begin{figure}
    \centering
    \includegraphics[width=\hsize]{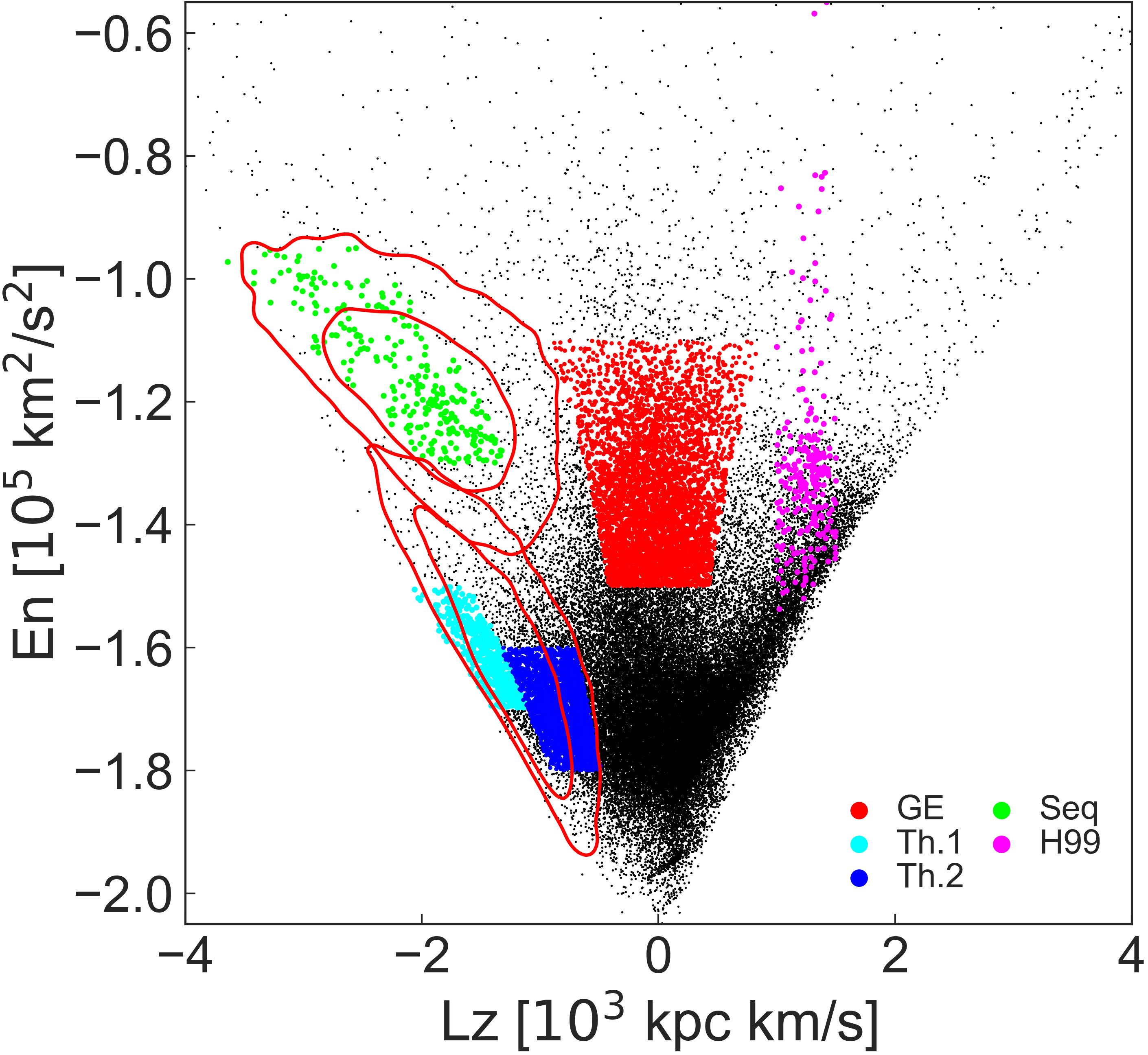}
    \caption{Distribution of stars in $En-L_z$ space, where the contours indicate the extent of mock dwarf galaxies placed on top of the debris of Sequoia and Thamnos 1 \& 2. The inner contour is for a dwarf galaxy of $M_\star = 5\cdot10^6~{\rm M}_\odot$ and the outer for $M_\star = 10^8~{\rm M}_\odot$. In the background we show all stars with halo-like kinematics within 3 kpc with small dots, see Sect.~\ref{sec:data}, with those belonging to selected structures colour-coded as in Fig.~\ref{fig:clustering-results}.}
    \label{fig:ELcontour}
\end{figure}

\subsection{Notes on Sequoia}
Using a mixture of spectroscopic data, the {\it Stellar Abundances for Galactic Archeology} (SAGA) database \citep{Suda2008StellarStars},
\cite{Matsuno2019OriginHalo} have shown that the trend defined by
Sequoia members is slightly offset from that of Gaia-Enceladus. Despite 
that we only have three stars with full abundance information provided by
APOGEE, we seem to derive a similar conclusion.

In spite of using the same APOGEE dataset as \cite{Myeong2019EvidenceHalo}
we reach different conclusions on the nature of Sequoia. One of the
reasons driving this is that we find that it is more natural to
separate the very retrograde halo into a high-$En$ (Sequoia) and a
low-$En$ (Thamnos) substructures. Sequoia is not only chemically
different from Thamnos, but it seems also difficult to reconcile its metallicity with the
large extent in $En$ proposed by \citet[][starting from the more unbound
Gaia-Enceladus debris down to the energy of
$\omega$-Cen]{Myeong2019EvidenceHalo}. Such a
large range in energy can only be produced by a very massive object,
as illustrated in Fig.~\ref{fig:ELcontour} where we have overlaid
contours on the $En-L_z$ diagram using the extent of mock dwarf
galaxies. The outer contour corresponds to a mock dwarf galaxy of
$M_\star = 10^8~{\rm M}_\odot$ and the inner to
$M_\star = 5\cdot10^6~{\rm M}_\odot$ \citep[see for details on the
mock dwarfs Sect.~5 of][]{Koppelman2019CharacterizationDR2}. The mocks
are centred on an orbit that is chosen to be roughly in the centre of
the debris of Sequoia and Thamnos in this diagram (only slight shifts are
found when a different central orbit is chosen). The contours
encompass 80\% of the stars in the mock dwarfs. Therefore the extent of an object in $En-L_z$ space reflects - to
some degree - the initial mass and size of the
progenitor. Fig.~\ref{fig:ELcontour} evidences that the contours of
the most massive mock dwarfs for the Sequoia and Thamnos overlap, but
confirms our assessment that they are likely distinct systems.

On the other hand, having lower binding energy and more retrograde motion than the bulk of the debris
of Gaia-Enceladus and overlapping with its metal-poor tail, Sequoia could well be at least in part, debris from the outer regions
of Gaia-Enceladus \cite[see Fig.1 of][]{Helmi2018}, lost at early times. This analysis suggests that at best, we are dealing with a bonsai Sequoia.

\subsection{Thamnos}
We find evidence for one or two more distinct components in the local retrograde halo: Thamnos 1 \& 2. The debris of these objects is characterised by strong retrograde rotation and high binding energy. Especially the values of $En$ suggests that these structures may have been accreted a very long time ago. The distribution of these structures in $En-L_z$ space is compatible with them originating in the same dwarf galaxy, see Fig.~\ref{fig:ELcontour}. Figure~\ref{fig:clustering-results} (bottom,right) shows that neither $\omega$-Cen nor FSR 1758 fall inside the selection boxes for Thamnos. When comparing to the full catalogue of \cite{Massari2019TheClusters}, we find no globular clusters to fall inside the selection for Thamnos. Compared to Gaia-Enceladus, the chemical composition of Thamnos' stars are more metal-poor and significantly more $\alpha$-enhanced. As far as we can judge and given their similar abundances, Thamnos 1 \& 2 share the same progenitor whose stellar mass $M_\star \lesssim 5 \times 10^6~{\rm M}_\odot$.

\subsection{The chemically defined thick-disc}
Our analysis reveals the presence of stars from the thick-disc with
retrograde motions, identified chemically because they are metal-rich and
$\alpha$-enhanced. It will be interesting to study these stars
detailed chemical composition: they are amongst the oldest stars that formed
in the {\it in situ} disk of the Milky Way. Since they were present at the time of the
merging of Gaia-Enceladus (what explains their hot orbits) such a
study would allow the characterisation of the disk at $z \gtrsim 2$. Early 
attempts of such studies have dated the merger event of Gaia-Enceladus 
\citep{DiMatteo2018} and found an ultra metal-poor disc component \citep{Sestito2019TracingStars}.

\subsection{Final note}
The main conclusion of this work is that even with the excellent {\it
  Gaia} DR2 data, putting the shattered pieces together to reconstruct
history, as in true Galactic archaeology, remains challenging at present. 
The different substructures identified in dynamical
space show significant overlap. Chemical tagging helps with
disentangling, but the current sample of high-quality and reliable
abundances is too small to lead to firm conclusions. At this point 
we are in desperate need for high-quality spectroscopic
observations of the halo stars to supplement the {\it Gaia} data, as
fortunately planned for WEAVE \citep{Dalton2012WEAVE:Telescope} and 4MOST \citep{deJong20124MOST:Telescope}. Furthermore, there may be a significant gain in comparing
the detailed properties of the substructures to tailored
high-resolution hydrodynamical simulations of mergers of satellites
with Milky Way-like galaxies.

\begin{acknowledgements}
      HHK and AH acknowledge financial support from a Vici grant from NWO. The Flatiron Institute is supported by the Simons Foundation. This work has made use of data from the European Space Agency (ESA) mission Gaia (http://www.cosmos.esa.int/gaia), processed by the Gaia Data Processing and Analysis Consortium (DPAC, http://www.cosmos.esa.int/web/gaia/dpac/consortium). Funding for the DPAC has been provided by national institutions, in particular the institutions participating in the Gaia Multilateral Agreement. In the analysis, the following software packages have been used: {\tt vaex} \citep{Breddels2018}, {\tt numpy} \citep{VanDerWalt2011TheComputation}, {\tt matplotlib} \citep{Hunter2007Matplotlib:Environment}
, {\tt jupyter notebooks} \citep{Kluyver2016JupyterWorkflows}
\end{acknowledgements}

\bibliographystyle{aa} 
\bibliography{references} 


\end{document}